\documentclass[twocolumn,aps,pre,showpacs]{revtex4-1}
\usepackage{makeidx}
\usepackage{amsmath}
\usepackage{amssymb}
\usepackage{graphicx}
\usepackage{times}
\usepackage{graphics}
\usepackage{dcolumn}
\usepackage{bm}
\newcommand{\PathToFigures}{}
\def\mathbi#1{\textbf{\em #1}}
\def\mbf#1{\mathbf{#1}}

\begin{document}

\title{  Random field and random anisotropy
  $\mathbi{O(N)}$ spin systems with a free surface }
\author{Andrei A. Fedorenko}
\affiliation{ CNRS UMR5672 -- Laboratoire de Physique de l'Ecole Normale
Sup{\'e}rieure de Lyon, 46, All{\'e}e d'Italie, 69007 Lyon, France }

\date{\today}
\pacs{68.35.Rh, 64.60.ae, 75.30.Kz}

\begin{abstract}
We study the surface scaling behavior of a semi-infinite $d$-dimensional
$O(N)$ spin system in the presence of quenched random field
and random anisotropy disorders. It is known that above
the lower critical dimension  $d_{\mathrm{lc}}=4$ the infinite models
undergo a paramagnetic-ferromagnetic transition for $N>N_c$ ($N_c=2.835$ for
random field and $N_c=9.441$ for random anisotropy). For $N<N_c$ and
$d<d_{\mathrm{lc}}$ there exists a quasi-long-range
ordered phase with zero order parameter and a power-law decay of spin correlations.
Using functional renormalization group  we derive the surface scaling laws
which describe the ordinary surface transition for $d>d_{\mathrm{lc}}$  and
the long-range behavior of spin correlations near the surface in
the quasi-long-range ordered phase
for $d<d_{\mathrm{lc}}$. The corresponding surface exponents are calculated
to one-loop order. The obtained results can be applied to the surface
scaling of periodic elastic systems in disordered media and amorphous magnets.
\end{abstract}

\maketitle

\section{Introduction}
\label{sec:introduction}

The phase diagram and critical properties of spin systems with quenched
disorder attracted considerable interest for decades.
One usually distinguishes two types of quenched disorder:
(i) random-temperature like disorder corresponding to randomness coupled to
the local energy density as, for example, in diluted
ferromagnets  \cite{stichcombe-83};
(ii) random field like disorder corresponding to the case when the order
parameter couples to a random symmetry breaking field~\cite{imry75}.
The influence of random-temperature disorder is rather well understood.
There exist several powerful methods to study the
phase behavior and criticality such as perturbative renormalization group (RG).
The effect of the random field disorder being more profound
is much less studied.
The prominent example is the critical behavior of
the random field Ising model (RFIM) which complete understanding
is still lacking despite significant numerical, analytical and experimental
efforts \cite{nattermann98}. It has been found that the perturbative
calculations including  standard RG methods lead to incorrect results, in
particular, to the so-called  dimensional reduction (DR). Analysis of the Feynman
diagrams giving the leading
singularities \cite{young77} or using supersymmetry \cite{parisi79}
predicts that the critical behavior of the RFIM
in $d$ dimension is the same as that of the pure system in $d-2$
dimensions. Consequently, the lower critical dimension of the RFIM below
which there is no true long-range order is expected to be
$d^{\mathrm{DR}}_{\mathrm{lc}}=3$.
However, the simple Imry-Ma arguments show that the lower critical dimension of
the RFIM is in fact $d_{\mathrm{lc}}=2$~\cite{imry75}. The deviation
from the DR prediction is also confirmed by the high-temperature
expansion \cite{gofman96} and real space RG~\cite{fortin96}.
The failure of DR can be explained by complicated
energy landscape which renders the perturbation theory spoiled to all orders by
unphysical averaging over multiple minima and maxima.
The latter can be formulated in terms of supersymmetry or replica
symmetry breaking \cite{wiese05,mezard92}.
To overcome this obstacle one needs a non-perturbative method or correct
resumming  the perturbation theory.

The considerable progress has been achieved last years in studying the $O(N)$ models
in which disorder couples to the $N$-component order parameter
either linearly  as in the random field (RF) case
or bilinearly as in a random anisotropy (RA) system.
These models are relevant for diverse physical systems including amorphous
magnets~\cite{harris73}, diluted antiferromagnets in a uniform
external magnetic field~\cite{fishman79}, liquid crystals in porous
media~\cite{LQ,feldman04}, nematic elastomers~\cite{{fridrikh97}},
critical fluids in aerogels~\cite{He3,volovik06,fedorenko07},
vortices in type II superconductors~\cite{blatter94}, and
stochastic inflation in cosmology \cite{kuhnel08}.
Similar to the RFIM these models suffer of DR~\cite{young77,pelcovits79}.
It was shown that the expansion around the lower
critical dimension  of the the RF
$O(N)$ model $d_{\mathrm{lc}}=4$ generates an infinite number
of relevant operators which can be parameterized by a single
function~\cite{fisher85}.
However, the RG flow of this function
has no analytic fixed point (FP).
Only almost two decades later, being inspired by the progress in disordered elastic
systems~\cite{fisher86,nattermann92,chauve01,ledoussal02}, it was realized that
the scaling properties of the RF and RA systems are encoded in
a nonanalytic FP~\cite{feldman00}.  The non-analytic FPs
control the paramagnetic-ferromagnetic phase
transitions in the RF and RA $O(N)$ model and allow one to compute the
critical exponents within $\varepsilon=d-4$ expansion~\cite{feldman02}.
The obtained exponents are different from the DR prediction.
The FRG calculations have been extended to two-loop order
\cite{doussal06,tarjus06} and the effect of long-range disorder correlations
has been studied~\cite{fedorenko07,fedorenko06dw}.
Using developed in Ref.~\cite{tarjus-series} truncated exact FRG it was
argued that spontaneous breaking of the supersymmetry which leads
to a breakdown of DR, occurs only below a critical dimension
$d_{\mathrm{DR}}\approx5.1$~\cite{tarjus12}.

A more peculiar issue concerns the phase diagram of the RF and RA
models below $d_{\mathrm{lc}}$. It is known that for the RF model
and models with isotropic distributions of random anisotropies
 true long-range order is forbidden below $d_{\mathrm{lc}}=4$
(for anisotropic distributions,  long-range order can occur even
below $d_{\mathrm{lc}}$~\cite{dudka84}). Nevertheless,
quasi-long-range order (QLRO) with zero order parameter and an
infinite correlation length can persist even for
$d^{*}_{\mathrm{lc}}(N)<d<d_{\mathrm{lc}}$, where
$d^{*}_{\mathrm{lc}}(N)$ is the lower
critical dimension for the paramagnetic-QLRO transition.
For example, the Gaussian variational approximation
predicts that the vortex lattice in disordered type-II superconductors can form the
so-called Bragg glass exhibiting slow logarithmic growth of
displacements~\cite{bragg}. This system can be mapped onto the three dimensional
RF $O(2)$ model, in which the Bragg glass corresponds to the QLRO
phase. Indeed, for $N<N_c$ and $d<d_{\mathrm{lc}}$, the FRG
equations have attractive  FPs which describe the QLRO phases of RF
and RA models~\cite{feldman00}.
Despite that the question of the lower
critical dimension of the paramagnetic-QLRO transition is still
controversial. In order to study the transition
between the QLRO phase and the disordered phase using FRG,
one has to go beyond the one-loop approximation.
The truncated exact FRG \cite{tarjus-series} and the two-loop
FRG \cite{doussal06} performed using a double  expansion in
$\sqrt{|\varepsilon|}$ and $N-N_{\mathrm{c}}$ provide an additional
singly unstable FP which controls the transition.
Both methods give qualitatively similar pictures of the FRG flows:
the critical and attractive FPs merge in some dimension
$d^*_{\mathrm{lc}}(N)<d_{\mathrm{lc}}$ which is considered as the lower
critical dimension of the paramagnetic-QLRO transition.
For the RF $O(2)$ model, both methods give approximately
the same estimation
$d^*_{\mathrm{lc}}\approx 3.8(1)$, and thus, suggest that there is
no Bragg glass phase in $d=3$. However, one has
to take caution when extrapolating results obtained for small
$\sqrt{|\varepsilon|}$ and $N-N_{\mathrm{c}}$.  Moreover, in contrast
to the model of Refs.~\cite{feldman00} and \cite{doussal06} which
belongs to the so-called ``hard-spin'' models,
the system studied in Ref.~\cite{tarjus-series} corresponds  to
``soft spins''. They can belong to different
universality classes since the soft spin model
allows for topological defects which destroy the QLRO.

The real systems, usually, are finite and have boundaries which effect
is twofold: (i) the free energy of the system in addition to the bulk contribution
proportional to the volume acquires  a new term proportional
to the area of the surface; (ii) the presence of boundaries
breaks the translational invariance. In general, this can modify
the behavior in the boundary region extended in the bulk only over distances
of the order of the bulk correlation length. However, at the
bulk critical point or in the QLRO phase, the bulk correlation
length is infinite so that one can expect that the effect
of boundaries to be more pronounced. Indeed, the presence of the
boundaries introduces a whole set of critical exponents  describing
the scaling behavior at and close to the boundary at criticality \cite{binder83}.
Several different classes of the surface transitions are known depending upon
boundary conditions \cite{lubensky75}.
The \textit{ordinary transition} corresponds to
the case when the surface magnetization is suppressed due to reduced
number of close neighbors near the boundary so that the surface
ordering is completely driven by the bulk magnetization.
If for some reason the coupling between spins on the surface is  sufficiently
enhanced with respect to the bulk coupling or there is an external
surface magnetic field,
the surface may order before
the bulk does. The latter is called the \textit{surface transition}. Then the system
can undergo the so-called \textit{extraordinary transition}
in the presence of ordered surface. The two lines of the extraordinary transition
and the surface transition meet at the multicritical point which is called the
\textit{special transition}. The last three transition can take place
only if the dimension of the surface $d-1$ is above the lower critical dimension
for the transition. These transitions have been studied for various systems
with discrete and continuous symmetries using different methods,
such as RG and numerical simulations (for review see
\cite{binder83,diehl86b,pleimling04}).

The effect of weak random temperature like disorder on the surface criticality
was studied using RG methods in Refs.~\cite{ohno92,usatenko01}.
However, not so much is known about the surface criticality in systems with
RF disorder. The phase diagram of the 3D semi-infinite RFIM  as a function
of the ratio of bulk and surface interactions and the ratio of bulk and surface
fields has been studied using a mean field approximation in Ref.~\cite{saber87}.
The surface criticality of the RFIM has been studied numerically
in Ref.~\cite{laurson05}. It was also shown that the RF disorder on
the surface of a 3D spin system with continuous symmetry destroys the long-range
order in the bulk, and, instead, a QLRO  emerges~\cite{feldman02-2}.
In this work we address the question how do the RF and RA disorder in the bulk
effect the behavior of spin systems with continuous symmetry in vicinity of
free surfaces. In particular, we consider the ordinary surface transition of the
RF and RA $O(N)$ models for $d>4$ and the spin correlations in the QLRO phase
near a free surface for $d<4$.

The paper is organized as follows. Section~\ref{sec:model} introduces the model.
In Sec.~\ref{sec:FRG} we renormalize the theory and derive the scaling laws.
In Sec.~\ref{sec:exponents} we
calculate the surface critical  exponents to one-loop order.
Section~\ref{sec:con} summarizes the obtained results.


\section{Model and scaling laws}
\label{sec:model}

We consider a $d$-dimensional semi-infinite $O(N)$ spin system which configuration
is given by the $N$-component  classical vector field $\mbf{s}( \mbf{r} )$
satisfying the fixed-length constraint $|\mbf{s}(\mbf{r})|^2=1$.
The position vector $\mbf{r}=(\mbf{x},z)$ has a $(d-1)$-dimensional component
$\mbf{x}$ parallel to the surface and a one-dimensional component $z\ge 0$
which is perpendicular to the surface $z=0$. It is convenient to introduce
short notations for the volume integral over half space
$\int_V:=\int_0^{\infty} dz \int d^{d-1} x$  and for the surface integral
$\int_S:= \int d^{d-1} x$.
The large-scale behavior of the disordered  spin system can be described by
the effective Hamiltonian
\begin{eqnarray}  \label{H}
\mathcal{H}\left[\mbf{s}\right] = \mathcal{H}_0\left[\mbf{s}\right]
+\mathcal{H}_{\mathrm{surf}}\left[\mbf{s}\right]+
\mathcal{H}_{\mathrm{dis}}\left[\mbf{s}\right],
\end{eqnarray}
consisting of the sum of three terms which result from the semi-infinite bulk,
surface and disorder in the bulk.
The contribution from the the semi-infinite bulk can be expressed
in the form of the well-known $O(N)$ nonlinear sigma model:
\begin{eqnarray} \label{H0}
 \mathcal{H}_0\left[\mbf{s}\right] &=& \int_V \left[
   \frac{1}{2} \left(\nabla \mbf{s}(\mbf{r})\right)^2
    - \mbf{h}\cdot \mbf{s}(\mbf{r})  \right],
\end{eqnarray}
where $\mathbf{h}$ is the magnetic field in the bulk. The surface
contribution to Hamiltonian can be written in its simplest form
as~\cite{diehl86}:
\begin{eqnarray} \label{Hsurf}
 \mathcal{H}_{\mathrm{surf}}\left[\mbf{s}\right] &=&
  - \int_S   \mbf{h}_1 \cdot \mbf{s}(\mbf{x}),
\end{eqnarray}
where for simplicity we assume that the surface magnetic field $\mbf{h}_1$
has the same direction as the bulk field $\mbf{h}$.
We consider a quite general type of bulk disorder such that its potential
can be expanded in spin variables as follows
\begin{equation} \label{V0}
\mathcal{H}_{\mathrm{dis}}\left[\mbf{s}\right]
= -\int_V\, \sum\limits_{\mu=1}^{\infty}\sum\limits_{i_1...i_ {\mu}}
h^{(\mu)}_{i_1...i_ {\mu}}(\mbf{r}) s_{i_1}(\mbf{r}) ... s_{i_{\mu}}(\mbf{r}).
\end{equation}
The coefficients $h^{(\mu)}_{i_1...i_ {\mu}}(\mbf{r})$ are
assumed to be Gaussian random variables with zero mean and variances given by
\begin{eqnarray}
\overline{h^{(\mu)}_{i_1...i_{\mu}}(\mbf{r}) h^{(\nu)}_{i_1...j_{\nu}}
(\mbf{r}')}& =&  \delta^{\mu\nu}\delta_{i_1 j_1}
... \delta_{i_{\mu} j_{\nu}} r_{\mu}\delta(\mbf{r}-\mbf{r}').
\end{eqnarray}
The first two coefficients have simple physical interpretation:
$h^{(1)}_i$ is a random magnetic field and $h^{(2)}_{ij}$ is a second-rank
random  anisotropy. The higher order coefficients
$h^{(\mu)}$ are higher order random anisotropies.
As was shown in Ref.~\onlinecite{fisher85},
even if the system has only finite number of nonzero bare $h^{(\mu)}$, the
RG transformations will generate an infinite set of higher-order
anisotropies. However, the RG flow preserves the symmetry with respect to
rotation $\mbf{s}\to - \mbf{s}$. For instance, starting from the bare
model with only a second-rank anisotropy only even-rank anisotropies will
be generated by the RG flow. We will reserve the notation RA for the systems  which
possess this symmetry and the notation RF for the systems which do not.

We employ the replica trick to average over disorder. Introducing $n$ replicas
of the original system and averaging their joint partition function over disorder
we obtain the replicated Hamiltonian
\begin{eqnarray}
&& \!\!\!\!\! \mathcal{H}_n =
    \int_V \left\{\sum_{a=1}^n \left[
   \frac{1}{2}  \left(\nabla \mbf{s}_a(\mbf{r})\right)^2
     - \mbf{h} \cdot \mbf{s}_a(\mbf{r}) \right]
     \right.
     \nonumber \\
&& \ \ \
\left. - \frac{1}{2T} \sum_{a,b=1}^n
  \mathcal{R}\left(\mbf{s}_{a}( r )\cdot\mbf{s}_{b}( \mbf{r} )\right)
    \right\}
         - \sum_{a=1}^n \int_S \ \mbf{h}_1 \cdot \mbf{s}_a(\mbf{x})
    ,\ \ \ \ \    \label{Hn}
\end{eqnarray}
where we have defined the function $\mathcal{R}(z)=\sum_{\mu}r_{\mu} z^{\mu}$.
The properties of the original disordered system  (\ref{H})
can be extracted in the limit $n\to 0$. According to the above definition of
the RF and RA models, the function $\mathcal{R}(z)$ is arbitrary in the
case of the RF model and even for the RA systems.

Power counting shows that $d_{\mathrm{lc}}=4$ is the lower critical
dimension of the model (\ref{Hn}) ~\cite{pelcovits79}. Above the
lower critical dimension
the RF and RA systems undergo a paramagnetic-ferromagnetic transition.
The scaling behavior at criticality is
controlled by a zero temperature fixed point (FP) similar to the
RFIM~\cite{bray85}, reflecting the fact
that disorder dominates over the thermal fluctuations. However, the
temperature is dangerously irrelevant. For instance, this
results in violation of the usual hyperscaling relation
and appearance of an additional universal exponent $\theta$ that
modifies the hyperscaling relation to \cite{nattermann98}:
\begin{eqnarray}
\nu(d-\theta) = 2-\alpha, \label{eq:hyper}
\end{eqnarray}
where $\nu$ and $\alpha$ are the correlation length and the specific heat
exponents. One also expects a dramatic slowing down as the transition is
approached with the characteristic relaxation time
$\ln \tau \sim  t_1 ^{-\nu \theta } $, where $t_1=|T-T_c|/T_c$
is the reduced temperature \cite{fisher86rf}.
The magnetization in the bulk and on the surface
vanish at the transition according to
\begin{eqnarray}
\sigma(t_1)\sim t_1^{\beta}, \  \  \   \   \ \sigma_1(t_1)\sim t_1^{\beta_1},
\label{eq:beta-exp-def}
\end{eqnarray}
where we have introduced the bulk and the surface magnetization exponents.
At the critical point $t_1=0$ a small magnetic field in the bulk $\mathbf{h}$
induces the magnetization in the bulk and also on the surface according to
\begin{eqnarray}
\sigma(h)\sim h^{1/\delta}, \  \  \   \   \ \sigma_1(h)\sim h^{1/\delta_1},
\label{eq:delta-exp-def-1}
\end{eqnarray}
where we define the exponents $\delta$ and $\delta_1$.
The surface magnetic field $\mathbf{h}_1$ leads to the surface magnetization
\begin{eqnarray}
\sigma_1(h_1)\sim h_1^{1/\delta_{11}}. \label{eq:delta-exp-def-2}
\end{eqnarray}
Below the lower critical dimension $d_{\mathrm{lc}}$ a QLRO phase
with zero magnetization can emerge.
At criticality  or in the QLRO phase, the correlation
functions of the order parameter exhibit scaling behavior.
Due to dangerous irrelevance of the temperature the connected and disconnected
correlation functions scale with different exponents.
We define the connected and disconnected correlation
functions of the two local operators $A$ and~$B$ as
\begin{eqnarray}
&&[A(\mbf{r})\cdot  B(\mbf{r}')]_{\mathrm{con}} :=
\overline{\langle A(\mbf{r})\cdot B (\mbf{r}') \rangle
 -   \langle A(\mbf{r})\rangle\cdot \langle B(\mbf{r}')\rangle},\ \ \
 \nonumber \\ 
&& [A(\mbf{r})\cdot B(\mbf{r}')]_{\mathrm{dis}} :=
\overline{\langle A(\mbf{r})\rangle
  \cdot \langle B(\mbf{r}')}\rangle -
\overline{\langle A(\mbf{r})\rangle}\cdot
\overline{ \langle B(\mbf{r}') \rangle}. \ \ \ \ \nonumber
\end{eqnarray}
Here the angular brackets denote the thermal averaging and the bar stands
for the disorder averaging.
For instance, the connected and disconnected correlation functions
of spins in the bulk scale independently as
\begin{eqnarray}
&& [\mbf{s}(\mbf{r}) \cdot  \mbf{s}(\mbf{r}')]_{\mathrm{con}} \sim
 \frac1{ |\mbf{r}-\mbf{r}'|^{d-2+\eta}},\label{eq:cor-fun-con} \\
&&  [\mbf{s}(\mbf{r}) \cdot  \mbf{s}(\mbf{r}')]_{\mathrm{dis}} \sim
\frac1{ |\mbf{r}-\mbf{r}'|^{d-4+\bar{\eta}}}. \label{eq:cor-fun-dis}
\end{eqnarray}
Following the general scaling picture of the surface critical phenomena
we introduce the surface exponents $\eta_{\bot}$ and $\bar{\eta}_{\bot}$
which replace the bulk exponents $\eta$ and $\bar{\eta}$
in Eqs.~(\ref{eq:cor-fun-con}) and
(\ref{eq:cor-fun-dis}) when one of the points $\mbf{r}$ or $\mbf{r}'$
belongs to the surface:
\begin{eqnarray}
&& [\mbf{s}(\mbf{x},z) \cdot  \mbf{s}(\mbf{x}',0)]_{\mathrm{con}} \sim
 \frac1{ \left((\mbf{x}-\mbf{x}')^2+z^2\right)^{(d-2+\eta_{\bot})/2}},\ \ \
 \label{eq:cor-fun-con-perp} \\
&&  [\mbf{s}(\mbf{x},z) \cdot  \mbf{s}(\mbf{x}',0)]_{\mathrm{dis}} \sim
\frac1{ \left((\mbf{x}-\mbf{x}')^2+z^2\right)^{(d-4+\bar{\eta})/2}}. \ \ \
\label{eq:cor-fun-dis-perp}
\end{eqnarray}
We also define the surface exponents
$\eta_{\parallel}$ and
$\bar{\eta}_{\parallel}$ which describe the connected and disconnected
correlation function when the both points lie on the surface:
\begin{eqnarray}
&& [\mbf{s}(\mbf{x}) \cdot  \mbf{s}(\mbf{x}')]_{\mathrm{con}} \sim
 \frac1{ |\mbf{x}-\mbf{x}'|^{d-2+\eta_{\parallel}}},\label{eq:cor-fun-con-paral} \\
&&  [\mbf{s}(\mbf{x}) \cdot  \mbf{s}(\mbf{x}')]_{\mathrm{dis}} \sim
\frac1{ |\mbf{x}-\mbf{x}'|^{d-4+\bar{\eta}_{\parallel}}}.
\label{eq:cor-fun-dis-paral}
\end{eqnarray}

Schwartz and Soffer~\cite{schwartz85} showed that the bulk exponents of the
RF model obey the inequality $2\eta\ge\bar{\eta}$.
The same arguments can be also applied to the surface correlation
functions so that the surface exponents
satisfy similar inequalities:
$2\eta_{\bot}\ge\bar{\eta}_{\bot}$  and
$2\eta_{\parallel}\ge\bar{\eta}_{\parallel}$.
Note, that these inequality cannot be applied to
the RA model where the coupling to disorder is bilinear.

\section{Functional renormalization group }
\label{sec:FRG}

\subsection{Perturbation theory}

In the limit of low temperature and weak disorder the configuration of the system is
fluctuating around  the completely ordered state
in which all replicas of all spins align along the same direction which is parallel
to  $\mbf{h}$ and $\mbf{h}_1$.
It is convenient to split the order parameter $\mbf{s}_a=({\sigma}_a,\bm{\pi}_a)$
into the $(N-1)$-component vector $\bm{\pi}_a$ which is perpendicular
to this direction  and the component ${\sigma}_a=\sqrt{1-{\bm{\pi}}_a^2}$
parallel to this direction.
Then the effective action of the system can be written as
\begin{eqnarray}
&& \!\!\! \mathcal{S}[\bm{\pi}]=
   \frac{1}{T} \sum_{a=1}^n \left\{  \int_V  \left[
   \frac{1}{2}  \left(\nabla \bm{\pi}_a \right)^2
   +\frac{(\bm{\pi}_a \cdot \nabla \bm{\pi}_a)^2}{2(1-\bm{\pi}_a^2)}
     - h\, \sigma_a \right]
     \right.
     \nonumber \\
&& \hspace{2mm}  - \left. \int_S \ h_1\, \sigma_a  \right\}
  - \frac{1}{2T^2} \sum_{a,b=1}^n \int_V
  \mathcal{R}\left( \bm{\pi}_a\cdot \bm{\pi}_b + \sigma_a \sigma_b
  \right).
     \label{eq:S-1}
\end{eqnarray}
In general one has to add to the action (\ref{eq:S-1}) the terms like
$\delta^d(0)\int_V \ln(1-\bm{\pi}_a^2)$ generated by
the Jacobian of the transformation from $\mathbf{s}_a$ to $\bm\pi_a$.
However, in what follows we will use the dimensional regularization
scheme\cite{zinn-justin} in which $\delta^d(0)=0$ so that
we ignore these terms in action~(\ref{eq:S-1}) from the beginning.

Let us denote averaging with the action~(\ref{eq:S-1}) by
double angular brackets and introduce the following correlation functions
\begin{eqnarray}\!\!\!\!\!\!
G^{(L,K)}_{\alpha,\beta}(\mbf{r},\mbf{x})=
{\left\langle\! \! \! \left\langle \prod_{\nu=1}^{L}
 {\pi}_{\alpha_{\nu}}(\mbf{r}_{\nu}) \prod_{\mu=1}^{K} {\pi}_{\beta_{\mu}}
(\mbf{x}_{\mu})\right\rangle\! \! \!\right\rangle}, \label{eq:cor-fun-2}
\end{eqnarray}
where $L$ points $\mathbf{r}=(\mathbf{r}_1,...,\mathbf{r}_L)$ are off surface
and $K$ points $\mathbf{x}=(\mathbf{x}_1,...,\mathbf{x}_K)$ are siting on the surface.
In Eq.~(\ref{eq:cor-fun-2}) we have used a short notation
$\alpha=(\alpha_1,...,\alpha_L)$ where each $\alpha_{\nu}$ stands for the component
number $i_{\nu}$ and the replica number $a_{\nu}$. The similar holds for $\beta$.
The correlation functions (\ref{eq:cor-fun-2}) can be computed using the following
generating functional~\cite{diehl83}
\begin{eqnarray}
 \mathcal{F}[\mathbf{J}, \mathbf{J}_1] =
  \ln
   \int \mathcal{D} \bm{\pi} e^{-\mathcal{S}[\bm{\pi}]+
  \int_V \mathbf{J}(\mathbf{r})\bm{\pi}(\mathbf{r})
  +\int_S \mathbf{J}_1(\mathbf{x})\bm{\pi}(\mathbf{x})}, \ \ \ \
  \label{eq:gen-fun-1}
\end{eqnarray}
where we assume that the source $\mathbf{J}(\mathbf{r})$ vanishes
at the surface. Differentiating with respect to the sources  we obtain
\begin{eqnarray}
G^{(L,K)}(\mbf{r},\mbf{x})= \left.
\prod_{\nu=1}^{L}\frac{\delta}{\delta{J}(\mbf{r}_{\nu})}
\prod_{\mu=1}^{K} \frac{\delta}{\delta {J}_{1}(\mbf{x}_{\mu})}
\mathcal{F}\right|_{J=J_1=0}, \ \
\end{eqnarray}
where for the sake of brevity we have suppressed all tensorial indices.
Using correlation
functions (\ref{eq:cor-fun-2}) one can compute the connected and disconnected
functions defined in Eqs.~(\ref{eq:cor-fun-con})  and (\ref{eq:cor-fun-dis}).
However, since we are interested only in the scaling behavior it is more convenient
to consider the similar correlation functions not for $\mathbf{s}$ but for $\bm\pi$
fields. For example, the correlation functions at two off surface
points read
\begin{eqnarray}\!\!\!\!\!\!
&& [\bm{\pi}(\mbf{r}) \cdot  \bm{\pi}(\mbf{r}')]_{\mathrm{con}} =
\lim\limits_{n\to 0} \sum\limits_{i=1}^{N-1} G^{(2,0)}_{i, a; i, a}
(\mbf{r},\mbf{r'}),
\label{eq:cor-fun-con-1} \\
&&  [\bm{\pi}(\mbf{r}) \cdot  \bm{\pi}(\mbf{r}')]_{\mathrm{dis}} =
\lim\limits_{n\to 0} \sum\limits_{i=1}^{N-1} G^{(2,0)}_{i, a; i, b}
(\mbf{r},\mbf{r'}).
\label{eq:cor-fun-dis-1}
\end{eqnarray}
where the connected correlator corresponds to a single replica and the disconnected
one to two different replicas $a\neq b$. To compute the correlation functions
at the surface like $[\bm{\pi}(\mbf{r}) \cdot  \bm{\pi}(\mbf{x}')]_{\mathrm{con}}$
or $[\bm{\pi}(\mbf{x}) \cdot  \bm{\pi}(\mbf{x}')]_{\mathrm{dis}}$ one has to replace
$G^{(2,0)}$ by $G^{(1,1)}$ and $G^{(0,2)}$, respectively.

Expanding the effective action (\ref{eq:S-1}) in small $\bm{\pi}$ we will
treat the quadratic part as a free action and the rest infinite series
as interaction vertices (see Appendix~\ref{sec:append-G}). Then the
correlation functions (\ref{eq:cor-fun-2}) can be expressed in terms of Feynman
diagrams which give the low temperature and small disorder expansion.
In practical calculations  it is convenient to perform the Fourier
transform with respect to $\mathbf{x}$:
$\hat{\bm{\pi}}(\mathbf{q},z)= \int d^{d-1} x \bm{\pi}(\mathbf{x},z)%
e^{-i \mathbf{q}\cdot \mathbf{x}}$ and define
$\int_q:= \int d^{d-1} q/(2\pi)^{d-1}$.
The quadratic terms give the free propagator
\begin{eqnarray}
\hat{G}_q^{(0)}(z,z')=\frac1{2 \bar{q}} \left[
e^{-\bar{q}|z-z'|}+\frac{\bar{q}+h_1}{\bar{q}-h_1}e^{-\bar{q}(z+z')}
\right], \label{eq:propagator-1}
\end{eqnarray}
where we have introduced the short notation $\bar{q}:= (q^2+h)^{1/2}$. The free
propagator~(\ref{eq:propagator-1}) satisfies the boundary conditions
\begin{equation}
\left.[\partial_z-h_1] G^{(0)}(\mathbf{x},z,\mathbf{x}',z')\right|_{z=0}=0.
\end{equation}
The free surface corresponds to the limit $h_1\to 0$ in which
Eq.~(\ref{eq:propagator-1}) becomes the Neumann propagator consisting of the
bulk part and the image part. In what follows we will use
the Neumann propagator as the bare one and treat
the terms proportional to $h_1$ as soft insertions~\cite{diehl86,fedorenko06}.

\subsection{ FRG equations and critical exponents}

The correlation functions~(\ref{eq:cor-fun-2}) calculated perturbatively
in small disorder and temperature
suffer of the UV divergences. To avoid mixture with IR singularities in the
$O(N)$-noninvariant correlation functions it is convenient to keep $\mathbf{h}\neq 0$.
The UV divergences can be converted into poles in $\varepsilon=d-4$ using
dimensional regularization.
To renormalize the theory one has to absorb these poles into finite number of
$Z$-factors. However, all the Taylor coefficients $r_{\mu}$ of the disorder
correlator $\mathcal{R}(\phi)$ turn out to be relevant operators so that one has
to introduce renormalization of the whole function. To simplify calculation
of the disorder
renormalization one can use the background field method~\cite{ledoussal02}.
Using the Legendre transform of the generating functional (\ref{eq:gen-fun-1})
from the sources $\mathbf{J}$ to the background fields $\bm\Pi$ one derives
the effective action $\Gamma[{\bm\Pi}]$  which is the generating functional
of the one-particle irreducible vertices. The two-replica part of the
effective action gives the  renormalization of the disorder.
Since the scaling behavior is controlled by a zero temperature
FP we will disregard all terms involving
more than two replicas which are suppressed in the limit $T\to 0$.
The renormalization of the disorder simplifies by changing variables:
$\mathcal{R}(\phi)= R(z)$ where $z=\cos \phi$, for instance,
$\mathcal{R}'(1)=- R''(0)$.
In terms of the variable $\phi$, the function
$R(\phi)$ becomes periodic with the  period $2\pi$ in the  RF case and
with the period $\pi$ in the RA case.
The relation between  the renormalized and the bare correlation functions reads
\begin{eqnarray}
G^{(L,K)}(\mbf{r}; T, h, h_1, R, \mu ) &=&
  Z_{\pi}^{-(L+K)/2}Z_{1}^{-K/2} \nonumber \\
&&\times \mathring{G}^{(L,K)}(\mbf{r}; \mathring{T}, \mathring{h},
\mathring{h_1},\mathring{R}). \label{eq:G0} \ \ \
\end{eqnarray}
where circles denote the bare quantities and $\mu$ is an
arbitrary momentum scale.
UV  divergences are absorbed into $Z$-factors according to
\begin{eqnarray}
&& \mathring{\bm{\pi}} = Z_{\pi}^{1/2}\bm{\pi},\hspace{12mm}
     \left.\mathring{\bm{\pi}}\right|_s
     = (Z_{\pi}Z_1)^{1/2}\left. \bm{\pi}\right|_s, \label{eq:Z-1}\\
&& \mathring{h}=\mu^{2}Z_T Z_{\pi}^{-1/2} h, \ \ \ \
 \mathring{h}_1=\mu Z_T (Z_{\pi} Z_1)^{-1/2} h_1, \label{eq:Z-2} \\
&&  \mathring{T}=\mu^{2-d} Z_T T, \hspace{8mm}
  \,\mathring{R}=\mu^{4-d} K_d^{-1} Z_R[R], \label{eq:Z-3}
\end{eqnarray}%
where $(2\pi)^d K_d=2 \pi^{d/2}/\Gamma(d/2)$ is the surface
area of a $d$-dimensional  unit sphere and
$\Gamma(x)$ is the Euler gamma function.
$Z_R[R]$ in Eq.~(\ref{eq:Z-3}) is a functional
acting on the renormalized disorder correlator $R(\phi)$ which has
the following loop expansion:
\begin{eqnarray}\label{eq:Z-R}
Z_R[R]=R+\delta^{(1)}(R,R)+\delta^{(2)}(R,R,R)+...,
\end{eqnarray}
where $\delta^{(1)}(R,R)$ is bilinear in $R$ and proportional to $1/\varepsilon$,
while $\delta^{(2)}(R,R,R)$ is cubic in $R$ and contains terms of order
$1/\varepsilon$ and $1/\varepsilon^2$.
According  to Eq.~(\ref{eq:Z-1}) the surface field $\left. \bm{\pi}\right|_s$
renormalizes differently from the field $\bm{\pi}$ in the bulk.
The new factor $Z_1$ serves to cancel the additional UV divergences in
Feynman diagrams arising  from the  image part of the Neumann  propagator
$\hat{G}_q^{(0)}(z,z')$ for $z'\to 0$.
The renormalized theory
is not unique and depends on the scale $\mu$. Using this fact we will derive the
functional renormalization group (FRG) equation.

We now consider how the scaling behavior can be extracted from the renormalized
theory. Using independence of the bare theory on the momentum scale $\mu$
one can derive the flow equations for the renormalized correlation functions
differentiating the both sides of Eq.~(\ref{eq:G0}) with respect to $\mu$ at fixed
bare quantities. One finds that the renormalized correlation functions satisfy
the following FRG equation
\begin{eqnarray}
&&\left[\mu \partial_{\mu}+(d-2-\zeta_T)T\partial_T -
\zeta_{h} h\partial_h - \zeta_{h_1} {h_1}\partial_{h_1} + \frac{L}2\zeta_{\pi}
\right.\nonumber \\
&&
 \left. + \frac{K}2( \zeta_{\pi}+ \zeta_1) - \int d\phi\, \beta[R(\phi)]
\frac{\delta}{\delta R(\phi)}\right]\,G^{(L,K)}=0, \label{eq:FRG-1}
\end{eqnarray}
where the integral in the last line is taken over a period, i.e.,
$(0,\pi)$ for RA and  $(0,2\pi)$ for RF models and  we have introduced
the scaling functions:
\begin{eqnarray}
&&\zeta_i = \left. \mu\partial_{\mu} \ln Z_{i } \right|_0, \  \  (i=T,\pi,1),
 \label{eq:zeta-i}\\
&&\zeta_h = 2 + \zeta_T - {\zeta_{\pi}}/2, \  \  \ \label{eq:zeta-h} \\
&& \zeta_{h_1} = 1 + \zeta_T - ({\zeta_{\pi}+\zeta_1})/2, \label{eq:zeta-h1}\\
&&\beta[R] = - \left. \mu\partial_{\mu} R(\phi) \right|_0. \label{eq:beta-R}
\end{eqnarray}
Here the zero indicates that the derivatives are taken at fixed bare quantities.
Flow equations similar to Eq.~(\ref{eq:FRG-1})  hold  also for the correlation
functions in which some or all the fields ${\bm \pi}_a(\mathbf{r})$ are
replaced by $\sigma_a(\mathbf{r})$ and for other observables, e.g., the
correlation length and the magnetization \cite{zinn-justin}.

The long-distance physics can be obtained from the solution of the FRG
equation~(\ref{eq:FRG-1}) in the limit of $\mu\to 0$.
The renormalized disorder correlator and the temperature flow according to
\begin{eqnarray}
&&-\mu\partial_{\mu} R(\phi)=\beta[R],  \label{eq:flow-RG} \\
&&   - \mu \partial_{\mu} \ln T = 2-d + \zeta_T. \label{eq:T}
\end{eqnarray}
The scaling behavior is controlled by a zero temperature FP
$\beta[R^*] =0$ with $R^*$ of order $\varepsilon $ and $T^*=0$.
Indeed, according to Eq.~(\ref{eq:T}), the temperature is irrelevant, i.e.
it flows to $0$ in the limit $\mu\to 0$ for $d>2$ and for sufficiently
small $\zeta_T=O(R)$. Although one expects that $\zeta_T$ is small in the
vicinity of the FP, one has to take caution whether the zero temperature FP
survives in three dimensions where $\zeta_T\sim\varepsilon$
is negative \cite{feldman00}.
The stability of the FP can be checked by computing the eigenvalues of the
disorder flow equation~(\ref{eq:flow-RG}) linearized about the FP solution:
$R(\phi)= R^*(\phi)+\sum_i t_i \Psi_i(\phi)$.
Since one expects that for $d>4$ ($\varepsilon>0$) the
FP $R^*(\phi)$ describes the paramagnetic-ferromagnetic transition it has to be
unstable in a single direction $\Psi_1(\phi)$ with eigenvalue $\lambda_1>0$:
$\beta[R^*+t_1\Psi_1]=\lambda_1 t_1 \Psi_1+O(t_1^2)$. In vicinity of the zero
temperature FP which controls the paramagnetic-ferromagnetic  transition,
the FRG equation for the correlation length $\xi$ can be written as
\begin{eqnarray}
&&\left[\mu \partial_{\mu} - \lambda_1 t_1
\frac{\partial}{\partial t_1 }\right]\xi(\mu,t_1)=0. \label{eq:FRG-xi}
\end{eqnarray}
Dimensional analysis implies that $\xi(\mu,t_1)=\mu^{-1} \bar{\xi}(t_1)$. This
reduces Eq.~(\ref{eq:FRG-xi}) to an ordinary differential equation (ODE) which
solution is $\xi \sim \mu^{-1} t_1^{-1/\lambda_1}$. The latter describes
divergence of the correlation length on the critical line at zero temperature
when the strength of disorder approaches
the critical value \cite{bray85}. Assuming that along the transition line
at finite temperature
$t_1\sim T-T_c$ we find that the positive  eigenvalue $\lambda_1$
gives the critical exponent of the correlation
length $\nu=1/\lambda_1$.
For $d<4$ ($\varepsilon<0$)
the FP becomes stable and describes a QLRO phase. The fluctuations
exhibit power-law correlations in the whole QLRO phase  so that the correlation
length $\xi$ is always infinite down to the lower critical dimension of the
QLRO - paramagnetic transition.

Let us consider the solution of Eq.~(\ref{eq:FRG-1}) for the connected
two-point correlation functions. The dangerous irrelevance of the temperature
manifests itself in the fact that the connected (bulk or surface) two point
functions are proportional to $T$ in the low temperature limit. This is explicitly
shown in Appendix~\ref{sec:append-G} for the connected correlation function
$G^{(1,1)}$. Hence, setting $h=h_1=0$ and $R=R^*$ we can rewrite
Eq.~(\ref{eq:FRG-1}) as
\begin{eqnarray}
&&\left[\mu \partial_{\mu}
 + \frac12(L+K)\zeta_{\pi}^*  + \frac{K}2 \zeta_1^* + \theta \right]
 \,G^{(L,K)}_{\mathrm{con}}=0, \label{eq:FRG-2}
\end{eqnarray}
where the star denotes that the function is computed at the FP.
In Eq.~(\ref{eq:FRG-2}) we have defined the exponent
\begin{eqnarray}
\theta=d-2-\zeta_T^*,
\end{eqnarray}
which describes the flow of the temperature (\ref{eq:T}) in the vicinity of the FP
and which has been introduced  ad hoc in the modified  hyperscaling
relation~(\ref{eq:hyper}).
Using the method of characteristics and dimensional analysis one can
write the solution of Eq.~(\ref{eq:FRG-2}) in the form
\begin{eqnarray} \label{eq:Gcon-RG}
G^{(L,K)}_{\mathrm{con}}( r b; R^*) = b^{-(
 \frac12(L+K)\zeta_{\pi}^*  + {K}\zeta_1^*/2 +\theta)} f_c(r; R^*). \ \ \
\end{eqnarray}
Considering the connected two point functions (\ref{eq:Gcon-RG})
with $(L=2,K=0)$, $(L=1,K=1)$, and $(L=0,K=2)$ we derive the critical
exponents:
\begin{eqnarray}
&& \eta=\zeta_{\pi}^*-\zeta_{T}^*, \label{eq:eta-1}\\
&& \eta_{\perp}=\zeta_{\pi}^*+\zeta_{1}^*/2-\zeta_{T}^*, \label{eq:eta-2}\\
&& \eta_{\parallel}=\zeta_{\pi}^*+\zeta_{1}^*-\zeta_{T}^*.\label{eq:eta-3}
\end{eqnarray}

We next turn to the disconnected two-point correlation functions. At variance
with the connected correlation functions they are not proportional to the temperature.
Thus, at $h=h_1=T=0$ they satisfy the same Eq.~(\ref{eq:FRG-2}) but without
the term $\theta$ in the brackets. The solution of the latter FRG equation
is given by
\begin{eqnarray}
G^{(L,K)}_{\mathrm{dis}}( r b; R^*) = b^{-(
 \frac12(L+K)\zeta_{\pi}^*  + {K}\zeta_1^*/2)} f_d(r; R^*). \ \ \
\end{eqnarray}
Repeating analysis we did for the connected functions we arrive at
\begin{eqnarray}
&& \bar{\eta}=4-d+\zeta_{\pi}^*=2+\eta-\theta, \label{eq:eta-bar-1}\\
&& \bar{\eta}_{\perp}=4-d+\zeta_{\pi}^*+\zeta_{1}^*/2=2+\eta_{\perp}-\theta,
  \label{eq:eta-bar-2} \\
&& \bar{\eta}_{\parallel}=4-d+\zeta_{\pi}^*+\zeta_{1}^*=2+\eta_{\parallel}-\theta.
   \label{eq:eta-bar-3}
\end{eqnarray}
Note that the exponents (\ref{eq:eta-1})-(\ref{eq:eta-3})
and (\ref{eq:eta-bar-1})-(\ref{eq:eta-bar-3}) are related by
\begin{eqnarray}
2{\eta}_{\bot}={\eta}+{\eta}_{\parallel}, \  \  \  \
2\bar{\eta}_{\bot}=\bar{\eta}+ \bar{\eta}_{\parallel}.
\label{eq:eta-perp-rel}
\end{eqnarray}

Finally we study the profile of the spontaneous magnetization below
and at the paramagnetic-ferromagnetic transition for $d>d_{\mathrm{lc}}$.
The magnetization as a function of the distance to the surface $z$, the
reduced temperature $t_1$, and the bulk and surface magnetic fields $h$ and
$h_1$ satisfies the following flow equation
\begin{eqnarray}
&&\left[\mu\partial_{\mu} -
\zeta_{h}^* h\partial_h - \zeta_{h_1}^* {h_1}\partial_{h_1}\right. \nonumber \\
&&\ \ \ \ \left. + \frac12 \zeta_{\pi}^*  + \frac{j}2\zeta_1^*
- \lambda_1 t_1
\frac{\partial}{\partial t_1 }\right]\sigma(z,t_1,h,h_1)=0. \label{eq:FRG-sigma}
\end{eqnarray}
Here $j=0$ and $z>0$ corresponds to the bulk magnetization~$\sigma$ while
$j=1$ and $z=0$ gives the surface magnetization~$\sigma_1$. The solution
of Eq.~(\ref{eq:FRG-sigma}) can be written as
\begin{eqnarray}
\sigma(z,t_1,h,h_1)&=& b^{-(\frac12 \zeta_{\pi}^*  + \frac{j}2\zeta_1^*)}
  \nonumber \\
&& \times \sigma(z b^{-1},t_1 b^{\lambda_1},h b^{\zeta_{h}^*},
  h_1 b^{\zeta_{h_1}^*}). \label{eq:sigma-sol}
\end{eqnarray}
We first consider the profile for $h=h_1=0$.
The solution (\ref{eq:sigma-sol}) interpolates  between the surface
magnetization $\sigma_1(t_1)\sim t_1^{(\zeta_{\pi}^*  + \zeta_1^*)/(2\lambda_1)}$
at $z\approx0$ and the bulk magnetization
$\sigma(t_1,z)\sim t_1^{\zeta_{\pi}^* /(2\lambda_1)}$
for $z\gg\xi$. Reexpressing the latter in terms of $\nu$,$\bar{\eta}$, and
$\bar{\eta}_{\parallel}$ we obtain
that the bulk and the surface magnetization exponents defined in
Eq.~(\ref{eq:beta-exp-def})
are given by
\begin{eqnarray}
\beta = \frac12 \nu (d-4+\bar{\eta}), \ \ \ \
\beta_1 = \frac12 \nu (d-4+\bar{\eta}_{\parallel}). \label{eq:beta-exp}
\end{eqnarray}
At the critical point $t_1=0$ and finite external fields we find that
$\sigma(h)\sim h^{\zeta_{\pi}^* /(2\zeta_{h}^*)}$ in the bulk
and $\sigma_1(h)\sim h^{(\zeta_{\pi}^*  + \zeta_1^*)/(2\zeta_{h}^*)}$
or $\sigma_1(h_1)\sim h_1^{(\zeta_{\pi}^*  + \zeta_1^*)/(2\zeta_{h_1}^*)}$
at the surface. Thus, the exponents $\delta$, $\delta_1$, and $\delta_{11}$
defined in Eqs.~(\ref{eq:delta-exp-def-1}) and (\ref{eq:delta-exp-def-2})
satisfy the following scaling relations:
\begin{eqnarray}
\frac{\delta-1}{2-\eta}=\frac{\nu}{\beta}, \ \ \ \ \
\frac{\delta_1-\beta/\beta_1}{2-\eta}=\frac{\nu}{\beta_1}, \ \ \ \ \
\frac{\delta_{11}-1}{1-\eta_{\parallel}}=\frac{\nu}{\beta_1}. \ \ \
 \label{eq:delta-exp}
\end{eqnarray}

\begin{figure}
\includegraphics[width=8cm]{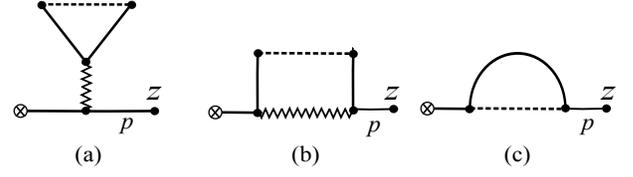}
\caption{One-loop diagrams contributing to the connected
two-point function $\mathring{G}^{(1,1)}_{1,a;1,a}(z,p)$. The solid
lines stand for the Neumann propagator ~(\ref{eq:propagator-1}). The wavy and dashed
lines are vertices defined in Eqs.~(\ref{eq:vertex-1})-(\ref{eq:vertex-3}). The
crossed circles denote the points on the surface.}
  \label{fig:diag}
\end{figure}


\section{The surface exponents to one-loop order}
\label{sec:exponents}

We now renormalize the both semi-infinite RF and RA models to one-loop order
and  explicitly calculate the surface critical exponents to first order in
$\varepsilon=d-4$.
The  factors $Z_{\pi}$, $Z_{T}$ and $Z_R[R]$ defined in
Eqs.~(\ref{eq:Z-1})-(\ref{eq:Z-R}) are the same that appear in the case of the
infinite systems. They have been calculated in several works
up to two-loop order \cite{fisher85,feldman00,feldman02,doussal06,tarjus06}.
To one-loop order they read
\begin{eqnarray}
&& Z_{\pi}=1-(N-1)\frac{R''(0)}{\varepsilon}+ O(R^2),\label{eq:Z} \\
&& Z_{T}=1-(N-2)\frac{R''(0)}{\varepsilon}+ O(R^2), \label{eq:ZT} \\
&& \varepsilon \delta^{(1)}(R,R) =  \frac12 R''(\phi)^2 -
   R'' (0) R'' (\phi)\nonumber \\
 && \hspace{10mm}  - (N-2) \bigg\{ R'' (0)[2  R (\phi) + R'(\phi)\cot\phi]
  \nonumber \\
 && \hspace{10mm} - \frac1{2\sin^2{\phi}} \big[ R'(\phi)  \big]^2 \bigg\}.
 \label{eq:ZR}
\end{eqnarray}
The new  factor $Z_{1}$ which eliminates the poles resulting from
the presence of the surface can be determined from
the renormalization of the two point function
$\mathring{G}^{(1,1)}(p,z;\mathring{h},\mathring{T},\mathring{R})$.
The one-loop diagrams contributing to this function are shown in
Fig.~\ref{fig:diag}. The corresponding integrals are computed in
Appendix~\ref{sec:append-G} and give
\begin{eqnarray}
&& \!\!\!\! \mathring{G}^{(1,1)}(p,z;\mathring{h},\mathring{T},\mathring{R})
 =\mathring{T} \frac{e^{-\bar{p} z }}{\bar{p}}\left\{1-
\frac{K_d}{4\varepsilon}\mathring{R}''(0)
 \right.
\nonumber \\
&& \left.
 \times \left[(N-3) \left(\frac{\mathring{h}}{\bar{p}^2}+\frac{z\mathring{h}}{\bar{p}}\right)
+2(N+1) \right]
+O(\mathring{R}^2) \right\},\ \ \ \ \label{eq:G11}
\end{eqnarray}
where $\bar{p}=(p^2+\mathring{h}^2)^{1/2}$.
The factor $Z_1$ can be found from the renormalization condition
\begin{eqnarray}
Z_{\pi}^{-1}Z_{1}^{-1/2}
\mathring{G}^{(1,1)}(p,z;\mathring{h},\mathring{T},\mathring{R}) =
\mathrm{finite\
for\ \ } \varepsilon \to 0,
\end{eqnarray}
where the bare $\mathring{h},\mathring{T},\mathring{R}$  are replaced by the
renormalized $h$, $T$ and $R$ according to Eqs.~(\ref{eq:Z-1})-(\ref{eq:Z-3}).
We obtain
\begin{eqnarray}
Z_{1}=1-(N-1)\frac{R''(0)}{\varepsilon}+ O(R^2).\label{eq:Z1}
\end{eqnarray}
Thus, to one loop order we have $Z_{1}=Z_{\pi}+O(R^2)$.
Using Eqs.~(\ref{eq:zeta-i}) and (\ref{eq:beta-R}) we calculate the
scaling functions
\begin{eqnarray}
\zeta_{T}&=&-(N-2){R''(0)}+ O(R^2), \label{eq:zeta-T-2} \\
\zeta_{\pi}&=&\zeta_{1}=-(N-1){R''(0)}+ O(R^2), \label{eq:zeta-pi-2}
\end{eqnarray}
and the beta function
\begin{eqnarray}
 \beta[R]
 &=& - \varepsilon R(\phi) + \frac12 R''(\phi)^2 - R'' (0) R'' (\phi)\nonumber \\
 &&  - (N-2) \bigg\{ R'' (0)[2  R (\phi) + R'(\phi)\cot\phi] \nonumber \\
 &&  - \frac1{2\sin^2{\phi}} \big[ R'(\phi)  \big]^2 \bigg\} + O(R^2)
 \label{eq:beta-1}
\end{eqnarray}
to one-loop order. Solution of the FP equation $\beta[R^*]=0$ with the beta
function~(\ref{eq:beta-1}) has been analyzed for different values of $N$
and different  sign of $\varepsilon$
in Refs.~\cite{feldman00,feldman02,doussal06,tarjus06}.
We first assume for granted that the flow has a FP $R^*(\phi)$ which is
a $\pi$-periodic function for the RA model and a $2\pi$-periodic function for the RF
model. Then, the surface critical exponents can be computed to one loop using
Eqs.~(\ref{eq:eta-1})-(\ref{eq:eta-3}) and
(\ref{eq:eta-bar-1})-(\ref{eq:eta-bar-3}) that yields
\begin{eqnarray}
&& \!\!\!\!\!\! \eta=-R^{*\prime\prime}(0), \ \ \ \ \ \ \ \ \ \ \ \ \ \ \ \ \
    \bar{\eta}=-\varepsilon-(N-1)R^{*\prime\prime}(0), \label{eq:eta-1r}\\
&&\!\!\!\!\!\! \eta_{\perp}=-\frac{N+1}2 R^{*\prime\prime}(0), \ \ \
   \bar{\eta}_{\perp}= -\varepsilon-\frac32(N-1)R^{*\prime\prime}(0),\ \ \
    \label{eq:eta-2r}\\
&&\!\!\!\!\!\! \eta_{\parallel}=-N R^{*\prime\prime}(0), \ \ \ \ \ \ \ \ \ \ \ \
   \bar{\eta}_{\parallel}=-\varepsilon-2(N-1)R^{*\prime\prime}(0). \ \
    \label{eq:eta-3r}
\end{eqnarray}
The other surface exponents are related to~(\ref{eq:eta-1r})-(\ref{eq:eta-3r})
by the scaling relations (\ref{eq:beta-exp}) and (\ref{eq:delta-exp}).

Before we explicitly calculate the surface exponents  for the semi-infinite
RF and RA models let us remind how
the FRG allows one to overcome the DR problem.
The incorrect DR prediction results from the assumption
that the flow equation (\ref{eq:flow-RG}) with the
beta function (\ref{eq:beta-1}) has a FP which is an analytic function.
Indeed, in this case one can obtain a closed flow equation for the
$R^{\prime\prime}(0)$:
\begin{eqnarray}
-\mu\partial_{\mu}R^{\prime\prime}(0)=-
 \varepsilon R^{\prime\prime}(0)- (N-2) R^{\prime\prime}(0)^2. \label{eq:flow-R2}
\end{eqnarray}
Equation~(\ref{eq:flow-R2}) has a nontrivial FP solution
$R^{*\prime\prime}(0)= - \varepsilon/(N-2)$ with the
eigenvalue $\lambda_1=\varepsilon$. This FP is unstable for $\varepsilon>0$
as one expects for a FP corresponding to the transition and gives the DR exponents:
$\nu^{(\mathrm{DR})}= 1/\varepsilon$ and
\begin{eqnarray}
&&\eta^{(\mathrm{DR})}=\bar{\eta}^{(\mathrm{DR})}=\frac{\varepsilon}{N-2}, \\
&&\eta^{(\mathrm{DR})}_{\perp}=\bar{\eta}^{(\mathrm{DR})}_{\perp}=
   \frac{N+1}{2(N-2)}\varepsilon, \\
&& \eta^{(\mathrm{DR})}_{\parallel}=\bar{\eta}^{(\mathrm{DR})}_{\parallel}
=\frac{N}{N-2}\varepsilon. \\
\end{eqnarray}
The one-loop DR exponents for the magnetization read
\begin{eqnarray}
\beta^{(\mathrm{DR})}=\frac{N-1}{2(N-2)}, \ \ \
\beta_1^{(\mathrm{DR})}=\frac{N-1}{N-2}.
\end{eqnarray}
For $\varepsilon<0$ the FP is stable but the $\eta$ critical exponents  become
negative, and hence, unphysical.

More accurate analysis of the RG flow shows
that $R'''(0)$ diverges at a finite scale $\mu$.
Thus, no analytic FP can exist and one has to look for a non-analytic FP with
$R^{*\prime\prime\prime}(0^+)\neq 0$ which would violate the DR predictions.
This requires solution of the boundary-value problem for the
nonlinear ODE with periodic boundary conditions, which
depend on the universality class. We assume that the
small $\phi$ expansion of the FP solution $R^*(\phi)$ has the following form
\begin{equation} \label{eq:r-phi}
R^*(\phi)=a_0+a_2 \phi^2 + a_3 |\phi|^3 + a_4 \phi^4 + a_5 |\phi|^5 +...,
\end{equation}
meaning that  $R^{*\prime\prime}(\phi)$ has a cusp at the origin with
$R^{*\prime\prime\prime}(0^+)\neq 0$.
Substituting ansatz (\ref{eq:r-phi}) into the FP equation we find that the
first coefficients are given by
\begin{eqnarray}
\!\!\!\!
&&a_0=-\frac{2 a_2^2 (N-1)}{4(N-2)a_2+\varepsilon}, \ \ \  \ \ \
 a_2= \frac{R^{*\prime\prime}(0)}2,
  \\
&&a_3= -\mathrm{sign}(\varepsilon)
  \sqrt{\frac{ 2\varepsilon a_2 + 4 a_2^2 (N-2) }{9(N+2)}}.\ \ \ \label{eq:FP-RF-A}
\end{eqnarray}
The value of  $R^{*\prime\prime}(0)$ as well as the sign
of $a_3$ are constrained by
the boundary conditions. $R^{*\prime\prime}(0)$ can be determined using the
shooting method to fulfill the appropriate periodicity requirement.

\subsection{Random field $O(N)$ model}

\subsubsection{Paramagnetic-ferromagnetic transition for $d>4$ ($\varepsilon>0$)}

\begin{figure}
\includegraphics[width=8cm]{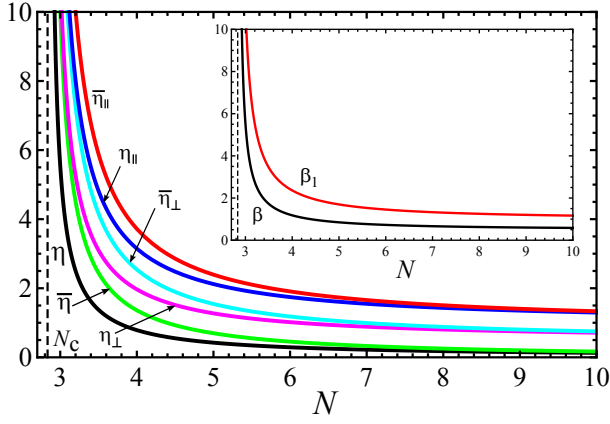}
\caption{(Color online) The critical exponents $\eta_i$ and
$\bar{\eta}_i$ (divided by $\varepsilon$), which
describe the paramagnetic-ferromagnetic transition of the RF model
above the lower critical dimension, as functions of $N$ for $N>N_c$.
Inset: The corresponding bulk magnetization exponent $\beta$ and the surface
magnetization exponent $\beta_1$ as functions of $N$. }
  \label{fig:2-rf-above}
\end{figure}

The RF model is described by $R(\phi)$ which is a $2\pi$-periodic function.
Numerical solution of the FP equation shows that for $d>4$
a $2\pi$-periodic solution of the
form (\ref{eq:r-phi})-(\ref{eq:FP-RF-A}) exists only for
$N>N_{\mathrm{c}}=2.834\,74$.
It has $R^{*\prime\prime}(0)<0$ and it disappears
when $N\to N_c^+$.
This cuspy FP is once unstable with the positive eigenvalue
$\lambda_1=\varepsilon$.
Thus, the correlation length exponent $\nu=1/\varepsilon + (\varepsilon^0)$
coincides with the DR prediction to one-loop order.
Remarkably, the non-zero $R^{*\prime\prime\prime}(0^+)$ vanishes
for $N>N^*=18+O(\varepsilon)$. The non-analyticity becomes weaker as $N$
increases and starts with $R^{*(2p(N)+1)}(0^+)\neq 0$ where $p\sim N$
\cite{tarjus06,doussal06,sakamoto06}.
Weaker non-analyticity  results in
restoring the DR critical exponents for $N>N^*$.
The critical exponents $\eta_i$ and $\bar{\eta}_i$
computed using Eqs.~(\ref{eq:eta-1r})-(\ref{eq:eta-3r})
as functions of $N$ are shown in Fig.~\ref{fig:2-rf-above}.  With
increasing $N$ they monotonically decay approaching the DR values at
$N=N^*$ and satisfying the inequalities:
$\eta<\bar{\eta}<\eta_{\perp}<\bar{\eta}_{\perp}<%
\eta_{\parallel}<\bar{\eta}_{\parallel}$.
The bulk and surface magnetization exponent $\beta$ and $\beta_1$
calculated for different $N$ are shown in inset of Fig.~\ref{fig:2-rf-above}.
To one-loop order
they obey relation $\beta_1=2\beta$. Up to now the both magnetization exponents
have been studied only for the 3D RFIM where numerical simulations give
$\beta=0.0017\pm0.005$ \cite{middleton02} and
$\beta_1=0.23\pm0.03$ \cite{laurson05}. Thus, the ratio $\beta_1/\beta$
for the RF $O(N)$ systems in $d>4$ is much smaller than for the 3D RFIM.

\subsubsection{Quasi-long-range order for $d<4$ ($\varepsilon<0$)}

\begin{figure}
\includegraphics[width=7.8cm]{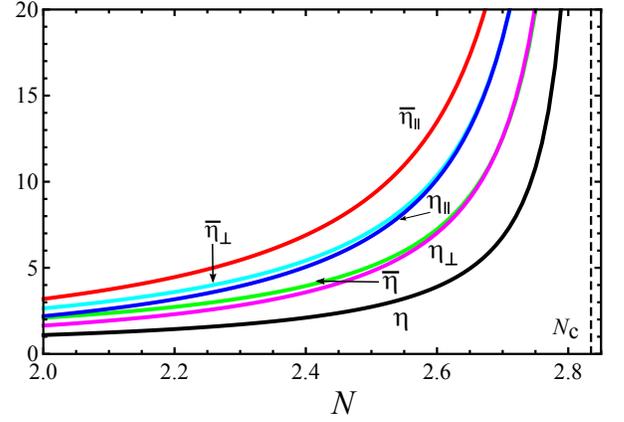}
\caption{(Color online) The critical exponents $\eta_i$ and
$\bar{\eta}_i$ (divided by $|\varepsilon|$), which describe the power-law
decay of correlations in the QLRO phase of the RF model below the lower critical
dimension, as functions of $N$ for $N<N_c$.
}
  \label{fig:3-rf-below}
\end{figure}

Below the lower critical dimension the flow equation for the disorder correlator
has an attractive $2\pi$-periodic FP solution of the form
(\ref{eq:r-phi})-(\ref{eq:FP-RF-A}).
This cuspy FP appears only for
$2\le N<N_c$ where it controls the scaling behavior of spin fluctuations in
the QLRO phase. The corresponding exponents
$\eta_i$ and  $\bar{\eta}_i$
as functions of $N$  are shown in Fig.~\ref{fig:3-rf-below}.
In the case $N=2$ the FP equation  admits for an explicit non-analytic
$\phi_0$-periodic  solution given by
\begin{equation} \label{eq:FP-periodic}
R^*(\phi)=\frac{|\varepsilon| \phi_0^4}{72}\left[\frac1{36}
-\left(\frac{\phi}{\phi_0}\right)^2\left(1-\frac{\phi}{\phi_0}\right)^2\right].
\end{equation}
Using Eqs.~(\ref{eq:eta-1r})-(\ref{eq:eta-3r}) one obtains
\begin{eqnarray}
&&\eta=\frac{\phi_0^2}{ 36} |\varepsilon|,\ \ \ \ \label{eq:eta-period-1}
   \bar{\eta}=\left(1+\frac{\phi_0^2}{ 36} \right) |\varepsilon|,  \\
&& \eta_{\perp} =\frac{\phi_0^2}{ 24}  |\varepsilon|, \ \ \ \ \label{eq:eta-period-2}
   \bar{\eta}_{\perp}=\left(1+\frac{\phi_0^2}{24} \right) |\varepsilon|, \\
&& \eta_{\parallel} =\frac{\phi_0^2}{18}  |\varepsilon|,  \ \ \ \
  \label{eq:eta-period-3}
\bar{\eta}_{\parallel}=\left(1+\frac{\phi_0^2}{ 18} \right) |\varepsilon|,
\end{eqnarray}
with $\phi_0=2\pi$ for the RF system.

The semi-infinite RF $O(2)$ model can be mapped onto a semi-infinite periodic
disordered elastic system with a free surface. There is one to one correspondence
between the Bragg glass phase of the elastic system and the
QLRO phase of the studied spin  model. The power-law decay of the spin
correlations in the QLRO phase corresponds to the logarithmic growth of the
displacements in the disordered elastic system.
Moreover, the exponents
$\eta$, $\eta_{\perp}$  and ${\eta}_{\parallel}$
provide the universal amplitudes of the logarithmic growth of the displacements
in the bulk, at the surface and along the surface, respectively.
For a $\phi_0$-periodic elastic system with a free surface these amplitudes
are given by Eqs.~(\ref{eq:eta-period-1})-(\ref{eq:eta-period-1}).
In particular,  we find that the logarithmic growth
of the displacements along the surface is twice larger than the logarithmic
growth in the bulk. In the case when only one point is on the surface
the growth is enhanced by $50\%$. The presence of a free surface can be considered
as an extended defect of a special kind.
The influence of potential-like extended defects on the Bragg-glass
has been recently studied in Refs.~\cite{fedorenko06dw,petcovic09}.

\subsection{Random anisotropy $O(N)$ model}

\subsubsection{Paramagnetic-ferromagnetic transition for $d>4$ ($\varepsilon>0$)}

\begin{figure}
\includegraphics[width=8cm]{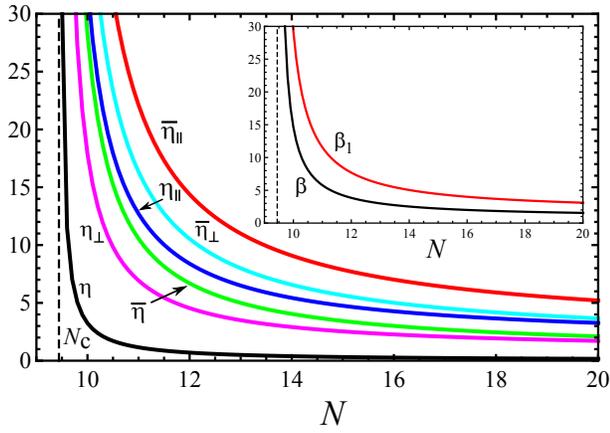}
\caption{(Color online) The critical exponents $\eta_i$ and
$\bar{\eta}_i$ (divided by $\varepsilon$), which
describe the paramagnetic-ferromagnetic transition of the RA model
above the lower critical dimension, as functions of $N$ for $N>N_c$.
Inset: The corresponding bulk magnetization exponent $\beta$ and the surface
magnetization exponent $\beta_1$ as functions of $N$. }
  \label{fig:4-ra-above}
\end{figure}

The FP equation has a cuspy $\pi$-periodic
solution of the form (\ref{eq:r-phi})-(\ref{eq:FP-RF-A})
which is singly unstable giving the correlation length
exponent $\nu=1/\varepsilon + (\varepsilon^0)$. It exists
for any $N>N_{\mathrm{c}}=9.4412$ with a non-zero
$R^{*\prime\prime\prime}(0^+)$. Therefore,
at variance with the RF case in the RA model the DR breaks down
for all values $N>N_c$, i.e., $N^*=\infty$ \cite{doussal06}.
The $N$-dependence of the critical exponents
$\eta_i$ and $\bar{\eta}_i$  is shown in
Fig.~\ref{fig:4-ra-above}.
For large $N$ one can find the asymptotic behavior of the FP
solution~\cite{tarjus06,doussal06,sakamoto06,doussal07reply}.
Following Ref.~\cite{tarjus06} we look for the
$\pi$-periodic solution of the FP equation $\beta[R]=0$ with the beta
function~(\ref{eq:beta-1}) of the form
\begin{eqnarray} \label{eq:ansatz-2}
R^{*\prime}(\phi)=-\frac32 \delta \varepsilon \sin\left(\frac{\pi-2\phi}{3}\right)
\left(2x(\phi)-1\right)G(x).
\end{eqnarray}
Here we have introduced a small parameter $\delta=1/(N-2)$ and defined variable
$x(\phi)=\cos(\frac{\pi-2\phi}{3})$. Substituting ansatz~(\ref{eq:ansatz-2})  into
the FP equation and expanding the function $G(x)$ in small $\delta$
one finds that the coefficients are polynomials in $x$:
\begin{eqnarray}
&&G(x)=1+\frac29 (95-44x-16x^2)\delta-
\frac{4}{81}(11737 - 5040 x \nonumber \\
&& - 3624 x^2 - 3104 x^3 -  768 x^4)\delta^2 + \frac{8}{10935}
(103378933 \nonumber \\
&&  - 45854072 x  - 23128624 x^2 -  16172328 x^3  \nonumber \\
&&- 9791216 x^4 - 4642048 x^5 -  901120 x^6)\delta^3 + O(\delta^4).
\end{eqnarray}
This implies that~\cite{doussal07reply}
\begin{eqnarray}
\!\!\!\!\!\!
\frac{R^{*\prime\prime}(0)} {\varepsilon\delta} = -\frac32 -
 23 \delta + \frac{1750}{3} \delta^2 - \frac{2129692}{27} \delta^3
  + O(\delta^4).\ \ \ \ \ \label{eq:R-ansatz}
\end{eqnarray}
Substituting the solution~(\ref{eq:R-ansatz}) into
Eqs.~(\ref{eq:eta-1r})-(\ref{eq:eta-3r})
 we find the correlation function exponents
to  leading order in $1/N$ as:
\begin{eqnarray}
\!\!\!\!\!
&&\eta=\frac{3\varepsilon}{2N}
\left(1+\frac{52}{3N}+...\right),\
\bar{\eta}=\frac{\varepsilon}{2}
\left(1+\frac{49}{N}+...\right), \ \ \\
&& \eta_{\perp}=\frac{3\varepsilon}{4}
\left(1+\frac{55}{3N}+...\right), \
\bar{\eta}_{\perp}=\frac{5\varepsilon}{4}
\left(1+\frac{147}{5N}+...\right), \ \ \ \ \ \ \ \ \  \\
&& \eta_{\parallel}=\frac{3\varepsilon}{2}
\left(1+\frac{52}{3N}+...\right), \
\bar{\eta}_{\parallel}={2\varepsilon}
\left(1+\frac{49}{2N}+...\right), \ \ \\
&& \beta= \frac{3}{4}
\left(1+\frac{49}{3N}+...\right), \ \ \ \
\beta_{1}=\frac{3}{2}
\left(1+\frac{49}{3N}+...\right),
\end{eqnarray}
where in the last line are the bulk and the surface magnetization exponents.

\subsubsection{Quasi-long-range order for $d<4$ ($\varepsilon<0$)}

\begin{figure}
\includegraphics[width=7.8cm]{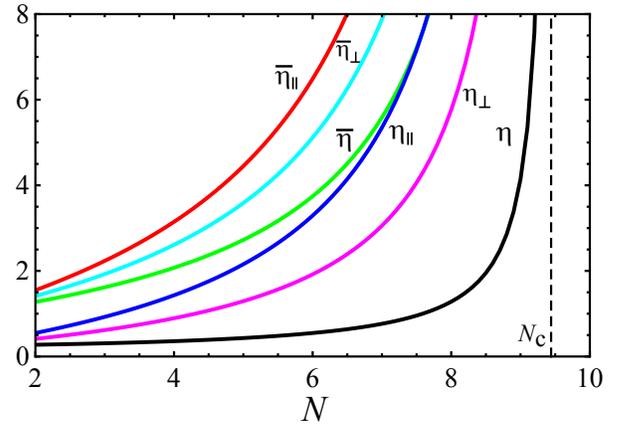}
\caption{(Color online) The critical exponents $\eta_i$ and
$\bar{\eta}_i$ (divided by $|\varepsilon|$), which describe the power-law
decay of correlations in the QLRO phase of the RA model below the lower critical
dimension, as functions of $N$ for $N<N_c$.
}
  \label{fig:5-ra-below}
\end{figure}

For $2\le N<N_c$ the flow equation has a stable $\pi$-periodic
FP solution of the form  (\ref{eq:r-phi})-(\ref{eq:FP-RF-A})
which controls
the scaling behavior of spin fluctuations in the QLRO phase of the RA model
for $d<4$.
The correlation function exponents $\eta_i$ and  $\bar{\eta}_i$
computed for different $N$
are shown in Fig.~\ref{fig:5-ra-below}.
For $N=2$ the FP equation has an explicit  non-analytic $\pi$-periodic
solution given by Eq.~(\ref{eq:FP-periodic}) with $\phi_0=\pi$.
The critical exponents of the RA $O(2)$ model are given by
Eqs.~(\ref{eq:eta-period-1})-(\ref{eq:eta-period-3}) with $\phi_0=\pi$.

\section{Summary}
\label{sec:con}

In the present work, we have investigated the RF and RA semi-infinite $O(N)$ models
with a free surface. The both models have the lower critical dimension
$d_{\mathrm{lc}}=4$.  Above $d_{\mathrm{lc}}$  they undergo a
paramagnetic-ferromagnetic transition for $N>N_c$, while below $d_{\mathrm{lc}}$
and for $N<N_c$ they exhibit a QLRO phase with zero magnetization and power-law
correlation of spins. Here the critical value $N_c=2.835$ for the RF models and
$N_c=9.441$ for the RA.
Using FRG we studied the surface scaling behavior
of these models at criticality as well as in the QLRO phase,
and calculate the corresponding surface exponents to lowest
order in $\varepsilon=d-4$. We have found that the DR prediction for the surface
scaling is broken similar to that happens in the bulk. We have shown
that the connected and disconnected correlation functions scale differently
also at the surface and derived the scaling relations between different surface
exponents.
The surface exponents obtained for the 3D RF $O(2)$
can be used to describe the growth of displacements near a free surface
in semi-infinite periodic elastic systems in disordered media.
The surface scaling we obtained
for the Heisenberg ($N=3$) RA model can be relevant for
the behavior of amorphous magnets \cite{harris73,amorphous}.

\acknowledgments

It is a pleasure to thank P.~Le~Doussal and K.~J.~Wiese
for stimulating discussions. I  would like to
acknowledge support from  the ANR grant 2010-Blanc IsoTop.

\appendix
\section{One-loop diagrams contributing to $\mathring{G}^{(1,1)}$  }
\label{sec:append-G}
In this appendix we calculate the correlation function
$\mathring{G}^{(1,1)}_{1,a;1,a}$ to one-loop order. Expanding action~(\ref{eq:S-1})
in small $\bm{\pi}_a$ we find
that the only vertices we need are
\begin{eqnarray}
&&\parbox{21mm}{\includegraphics[width=19mm]{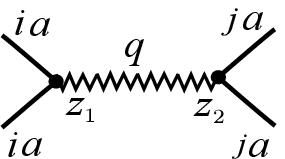}}
= - \frac1{8\mathring{T}} \delta(z_1-z_2)
 \left[q^2+\partial_{z_1}\partial_{z_2}+\mathring{h}\right],
\ \ \ \ \ \label{eq:vertex-1} \\
&&\parbox{21mm}{\includegraphics[width=19mm]{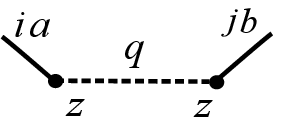}}
= - \frac1{2\mathring{T}^2} \mathring{R}''(0), \label{eq:vertex-2}\\
&&\parbox{21mm}{\includegraphics[width=19mm]{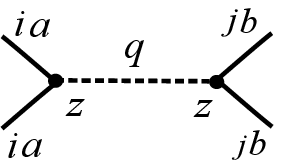}}
= - \frac1{8 \mathring{T}^2} \mathring{R}''(0). \label{eq:vertex-3}
\end{eqnarray}
The one loop diagrams contributing to the correlation function
$\mathring{G}^{(1,1)}_{1,a;1,a}$
are shown in Fig.~\ref{fig:diag}. The solid line
corresponds to the Neumann propagator (\ref{eq:propagator-1}) with $\mathbf{h}_1=0$
and the wavy and
dashed lines to vertices (\ref{eq:vertex-1})-(\ref{eq:vertex-3}). The first diagram
gives
\begin{eqnarray}
&& \!\!\!\!\!\! (a)=\frac{N-1}{2\mathring{T}^3}
  \mathring{R}''(0)\int_q\, \int\limits_{0}^{\infty} dz_1
\int\limits_{0}^{\infty} dz_2 \int\limits_{0}^{\infty} dz_3\, \delta(z_2-z_1)
\nonumber \\
&&\hspace{3mm} \times \left[\partial_{z_1}\partial_{z_2}+\mathring{h}\right]
G_p^{(0)}(0,z_1)G_p^{(0)}(z_1,z)\left[G_q^{(0)}(z_2,z_3)\right]^2 \nonumber \\
&&\hspace{3mm}  =\frac{N-1}{16\bar{p}} \mathring{T} \mathring{R}''(0) e^{-\bar{p} z }
\left( \frac{z \mathring{h}}{\bar{p}}+ \frac{\mathring{h}}{\bar{p}^2}
 + 2   \right)I_2 + \mathrm{finite},\ \ \ \ \
\end{eqnarray}
where we have used $\bar{p}=(p^2+\mathring{h}^2)^{1/2}$
and omitted the terms finite in the limit $\varepsilon\to 0$.
The logarithmically divergent one-loop integral reads
\begin{eqnarray}
I_2&=&\int_q \frac1{\bar{q} (\bar{p}+\bar{q} )^2} = K_{d-1} \int\limits_0^{\infty}
\frac{q^{d-2} d q }{(q^2+\mathring{h}^2)^{3/2}} + O(\varepsilon^0) \ \nonumber \\
&& = - \frac{4 K_d}{\varepsilon}+ O(\varepsilon^0).
\end{eqnarray}
The second and third diagrams yield
\begin{eqnarray}
&& \!\!\!\!\!\! (b)=\frac{1}{\mathring{T}^3} \mathring{R}''(0)\int_q\,
  \int\limits_{0}^{\infty} dz_1 \int\limits_{0}^{\infty} dz_2
   \int\limits_{0}^{\infty} dz_3\, \delta(z_2-z_1) \nonumber \\
&&\hspace{3mm}  \times \left[(\mathbf{p}+\mathbf{q})^2 + \partial_{z_1}
  \partial_{z_2}+\mathring{h}\right] G_p^{(0)}(0,z_1)  G_p^{(0)}(z_2,z) \nonumber \\
&&\hspace{3mm}  \times\,  G_q^{(0)}(z_1,z_3) G_q^{(0)}(z_3,z_2)
  = -\frac{1}{8\bar{p}} \mathring{T} \mathring{R}''(0) e^{-\bar{p} z }
  \nonumber \\
&&\hspace{3mm}   \times
\left[\left( \frac{z\mathring{h} }{\bar{p}} +\frac{\mathring{h}}{\bar{p}^2}-
 2 \bar{p} z - 7   \right)
 I_2 - 2I_3  \right] + \mathrm{finite},
\end{eqnarray}
\begin{eqnarray}
&& \!\!\!\!\!\! (c)=-\frac{1}{\mathring{T}^2} \mathring{R}''(0)\int_q\,
  \int\limits_{0}^{\infty}
   dz_1 G_p^{(0)}(0,z_1) G_q^{(0)}(z_1,z_1)  \, \nonumber \\
&&\hspace{3mm} \times G_p^{(0)}(z_1,z)
 =-\frac{1}{8\bar{p}} \mathring{T} \mathring{R}''(0) e^{-\bar{p} z }  \nonumber \\
&&\hspace{3mm} \times
\left[\left(2\bar{p} z +5   \right)I_2
+ 2I_3 \right] + \mathrm{finite},
\end{eqnarray}
where we have defined the algebraically divergent integral
\begin{eqnarray} \label{eq:divint}
I_3(\bar{p},z)=\int_q \frac{ 3 \bar{p}+\bar{q}+
 (2 \bar{p}+\bar{q})\bar{p} z}{  \bar{p}^2 (\bar{p} +\bar{q})^2}.
\end{eqnarray}
Summing up the three diagrams we find that the algebraically divergent
integral~(\ref{eq:divint}) cancels and we obtain Eq.~(\ref{eq:G11}).

\end{document}